# Shorter SPECT Scans Using Self-supervised Coordinate Learning to Synthesize Skipped Projection Views

Zongyu Li[1,2*], Yixuan Jia[1,2*†], Xiaojian Xu[1], Jason Hu[1], Jeffrey A. Fessler[1,2] and Yuni K. Dewaraja[2]

[1] Department of Electrical Engineering and Computer Science

[2] Department of Radiology, University of Michigan, Ann Arbor, MI, United States.

† Corresponding author: Yixuan Jia, Department of Electrical Engineering and Computer Science, University of Michigan, Ann Arbor, MI, United States, 48109-2122. jiayx@umich.edu; (01)734-846-2853

* First author:

Zongyu Li: zonyul@umich.edu; (01)734-223-1155.

Yixuan Jia: jiayx@umich.edu; (01)734-846-2853.

* Contributed equally to this work.

This research was supported by R01 EB022075 awarded by the National Institute of Biomedical Imaging and Bioengineering (NIBIB) and R01 CA240706 awarded by the National Cancer Institute (NCI), NIH

Running title: Shorter SPECT Scans with Self-supervised Coordinate Learning

Word count: 5568

# Abstract

**Purpose:** This study addresses the challenge of extended SPECT imaging duration under low-count conditions, as encountered in Lu-177 SPECT imaging, by developing a self-supervised learning approach to synthesize skipped SPECT projection views, thus shortening scan times in clinical settings.

**Methods:** We employed a self-supervised coordinate-based learning technique, adapting the neural radiance field (NeRF) concept in computer vision to synthesize under-sampled SPECT projection views. For each single scan, we used self-supervised coordinate learning to train a multi-layer perceptron (MLP) to estimate skipped SPECT projection views. The method was tested with various down-sampling factors (DFs=2, 4, 8) on both Lu-177 phantom SPECT/CT measurements and clinical SPECT/CT datasets, from 11 patients undergoing Lu-177 DOTATATE and 6 patients undergoing Lu-177 PSMA-617 radiopharmaceutical therapy. Performance was evaluated both in projection space and by comparing reconstructed images using 1) all measured views (ground-truth), 2) down-sampled (partial) measured views only, and partial measured views combined with skipped views 3) generated by linear interpolation and 4) synthesized by NeRF.

**Results:** NeRF-synthesized projections demonstrated lower Normalize Root Mean Squared Difference (NRMSD) relative to the measured projections. Averaged across the DFs, the NRMSD for NeRF-synthesized projections compared to linearly interpolated projections in the phantom study was 7% vs. 10%, in DOTATATE patient studies was 18% vs. 26% and in PSMA patient studies 20% vs. 21%. For SPECT reconstructions, NeRF outperformed the use of linearly interpolated projections and partial projection views in relative contrast-to-noise-ratios (RCNR) averaged across different DFs: 1) DOTATATE: 83% vs. 65% vs. 67% for lesions and 86% vs. 70% vs. 67% for kidney, 2) PSMA: 76% vs. 69% vs. 68% for lesions and 75% vs. 55% vs. 66% for organs, including kidneys, lacrimal glands, parotid glands, and submandibular glands.

**Conclusion:** The proposed method enables a significant reduction in acquisition time (by factors of 2, 4, or 8) while maintaining quantitative accuracy in clinical SPECT protocols by allowing for the collection of fewer projections. Importantly, the self-supervised nature of this NeRF-based approach eliminates the need for extensive training data, instead learning from each patient's projection data alone. The reduction in acquisition time is particularly relevant for imaging under low-count conditions and for protocols that require multiple-bed positions such as whole-body imaging.

# INTRODUCTION

SPECT/CT imaging has had many advances (*1*); however, one continuing limitation is that SPECT acquisition is slow, especially under the low-count conditions encountered when imaging therapy radionuclides, such as Y-90, Ac-225, Ra-223, and Lu-177. These radionuclides are chosen for the therapeutic properties of their alpha and beta emissions, hence do not have ideal properties for gamma-camera imaging. For example, the photon/gamma-ray yield is relatively low, leading to low count conditions. Nevertheless, it is very desirable to perform both therapy and imaging with the same radionuclide, even in very low-count applications.

With Lu-177 where the 208 keV gamma-ray intensity is only 10%, it can take 15-30 mins per bed (~40 cm axial) for SPECT on standard gamma-camera systems following radiopharmaceutical therapies (RPTs) such as Lu-177-DOTATATE and Lu-177-prostate-specific membrane antigen (PSMA) (*2,3*). For RPTs involving alpha-emitters, such as Ac-225-PSMA, acquisition times of up to 1 h have been proposed (*4*). This is because both the administered activities and the gamma-ray yields are very low. SPECT under low-count conditions is particularly challenging when multiple beds are needed to encompass metastases and critical organs throughout the body. For example, in PSMA therapy for metastatic castration-resistant prostate cancer (mCRPC), SPECT imaging may require 3-5 bed positions to include all critical organs such as lacrimal glands, salivary gland, bone marrow, and kidneys, as well as lesions that can be throughout the body (*5,6*). Such a procedure demands a significantly greater amount of camera time, which can not only lead to patient discomfort, but can also increase motion artifacts. Additionally, in many facilities, camera availability is limited (*6-10*).

To overcome these challenges, a shorter acquisition time is preferrable by taking either fewer projection views or shorter acquisition time per view. These strategies pose additional challenges due to either the missing (skipped) view angles or the increased image noise (*11*). Numerous algorithms have been proposed with a focus on denoising the reconstructed images from noisy projections to improve image quality (*12-23*). In contrast, the approach of synthesizing the missing projections (*24,25*) has been relatively unexplored.

Most prior studies have employed deep learning techniques to learn the relationship between one projection and its neighboring views, often relying on ground truth data for training purposes. For instance, Rydén et al. used a deep convolutional U-Net (*26*) trained to generate synthetic intermediate projections (*24*). Meanwhile, Li et al. introduced a network architecture called LU-Net that integrates Long Short-Term Memory network (*27*) and U-Net to understand the transformation from sparse-view projection data to full-view data (*28*). Chen et al. presented a cross-domain method using SPECT images predicted in the image domain as reference for synthesizing full-view projections in the sinogram domain (*29*). These approaches are reported to be effective, but they are all based on supervised learning methods that require a sufficient amount of paired data for training. However, in many cases, obtaining enough paired ground truth data for training is challenging or even infeasible. This difficulty is especially true in the case of post-therapy imaging for verifying uptake or dosimetry following RPT because such imaging is typically not part of routine clinical practice. On the other hand, self-supervised learning, which does not require separate training labels and instead learns from each scan itself, has the potential to overcome the limitations of supervised learning in such scenarios.

The aim of this research was to reduce SPECT acquisition time by reducing the required number of measured projection views while maintaining image quality by incorporating synthetic projections generated by deep neural networks. We implemented a multi-layer perceptron (MLP) and trained it to generate skipped SPECT projection views through self-supervised coordinate learning (*30*). We evaluated the performance of the proposed method both qualitatively and quantitatively in phantom studies and in patients imaged after Lu-177 DOTATATE therapy and Lu-177 PSMA therapy. This work was extended from an abstract submitted to the Society of Nuclear Medicine & Molecular Imaging (SNMMI) 2023 annual meeting (*31*).

# MATERIALS AND METHODS

## Phantom study

We used an elliptical phantom with six hot sphere inserts of volumes 2,4,8,16,30,114mL. These 'hot' spheres (having the same Lu-177 activity concentration of 0.22 MBq/mL) are placed in a 'warm' background (0.035 MBq/mL) to achieve a sphere-to-background ratio of 6.3:1. The sphere volumes of interest (VOIs), corresponding to the physical size, were defined on the CT images.

## Patient Studies

For the patient studies, we used SPECT/CT scan data from 11 patients imaged after Lu-177-DOTATATE therapy for neuroendocrine tumor (NET) and from 6 patients imaged after Lu-177-PSMA-617 therapy for mCRPC with the approval of University of Michigan Institutional Review Board (IRB) for retrospective analysis. We defined organs of interest (kidneys for DOTATATE therapy, and kidneys, lacrimal glands, parotid glands, and submandibular glands for PSMA therapy) using deep learning-based methods available within MIM Software®. A radiologist manually defined the lesions (78 in total, volume ranging from 2 to 250 mL) as described previously (*32*).

## SPECT/CT acquisition

All scans were acquired on a Siemens Intevo Bold SPECT/CT with a 5/8'' crystal equipped with medium-energy collimators. Acquisition parameters included 120 views, with 60 views per head, a 20% photopeak window centered at 208 keV, and two adjacent scatter windows of 10% width each. The phantom study used a prolonged acquisition of 196 sec/view to achieve count levels similar to that encountered in patient imaging after Lu-177 therapy. The patient images were acquired under the standard protocols used in our clinic. Lu-177-DOTATATE SPECT images were acquired for a single bed position at day 2 or day 4 after the cycle 1 administration of 7.4 GBq using an acquisition time of ~ 25 seconds per view (total scan time of ~ 25 min). TheLu-177-PSMA SPECT images were acquired with two bed positions at day 2 or day 3 after the cycle 1 administration of 7.4 GBq with an acquisition time of ~ 17 seconds per view per bed (total scan time of ~ 34 min). The projection view matrix size was 128×128, with a pixel size of 4.8×4.8mm. The CT images were acquired in low-dose mode (120 kVp; 15 – 80 mAs) under free breathing conditions, with a matrix size of 512×512 and pixel size of 0.98×0.98mm.

**Self-supervised Coordinate Learning**

Given the limited amount of data, we focused on a self-supervised learning approach, rather than supervised methods for this study. Our method draws inspiration from computer vision: the neural radiance field (NeRF) approach that models complex 3D scenes through a volumetric scene function (*33*). NeRF fundamentally uses neural networks to map 3D spatial coordinates to radiance values. In a similar vein, we developed a multi-layer perceptron (MLP), comprising 12 hidden layers with 256 neurons each, to synthesize missing projection views in SPECT imaging. Fig. 1 illustrates the input to our MLP: 5-dimensional coordinates for each pixel in SPECT projection views. These coordinates consist of pixel position (i, j), the sine and cosine of the view angle and radial position (to accommodate noncircular orbits). To enhance the representation of the continuous measurement field, we upscaled the original projection images by a factor of two with the nearest-neighbor resizing method. Consequently, the network input size for each projection view becomes (256×256)×5. The training target consists of measured counts, with a size of 256×256×1 for each view. During inference, the model is fed the coordinates of the missing SPECT projections and predicts the corresponding counts both for the main acquisition window and adjacent scatter windows. Our method provides flexible adaption to different numbers of projection views, corresponding to various down-sampling factors (DF). For instance, when trained on 30 measured views and synthesizing 90 views, it achieves a 75% reduction in scan time (DF=4). Additionally, in this study, we also tested our method for DF=2 and DF=8 cases.

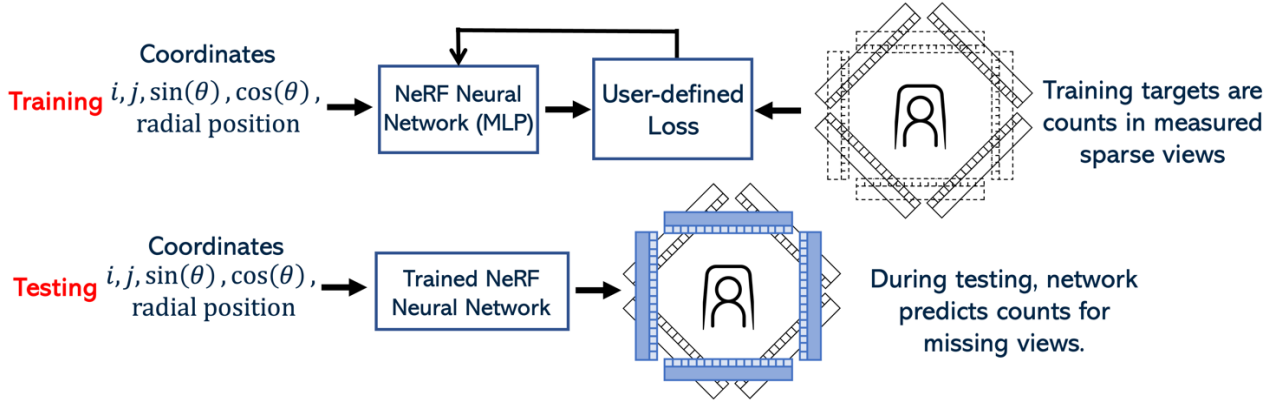

**Figure 1**. Workflow of the proposed SPECT projection synthesis method. The training process (top) involves inputting 5-dimensional coordinates into the MLP, with a user-defined loss function guiding the network to learn from the patient-specific training targets: measured counts in sparse views. During testing (bottom), the trained network receives the coordinates of missing views and outputs the predicted counts.

**Training and Optimization**

For each scan, we optimized the MLP weights by minimizing the Huber loss function ($\delta = 1$), which is less sensitive to outliers in data than the squared error loss (*34*), given as

$$L_\delta(a) = \begin{cases} \frac{1}{2}a^2 & \text{for } |a| < \delta \\ \delta \cdot (|a| - \frac{1}{2}\delta) & \text{otherwise} \end{cases}$$

quantifying the discrepancy between the model-predicted projections and the actual measurements. We employed the Adam optimizer (*35*) with an initial learning rate set at 0.001 and a reduce-on-plateau scheduler to minimize the loss function. We used coordinates corresponding to 20% of all pixels from the full projection views as per-patient validation data. The patient-specific model was selected at the lowest validation loss out of 200 training epochs. We used a batch size of 10,000 out of $256 \times 256 \times n_{\text{bed}} \times n_{\text{view}}$ projection pixel coordinates. The training/testing virtual patient phantoms images and the implementation of our method in PyTorch will be available at: https://github.com/ZongyuLi-umich/.

## SPECT Reconstruction

The ordered-subset expectation-maximization (OSEM) algorithm (*36*) is commonly used to reconstruct 3D SPECT images. It estimates the emission distribution $x$ from noisy measurement projection data $y$, often statistically modeled as: $y \sim \text{Poisson}(Ax + \bar{r})$, where $A$ characterizes the system matrix that incorporates factors such as attenuation and detector efficiency, and $\bar{r}$ denotes the mean of background events such as scatter events.

In this study we performed OSEM SPECT reconstructions (DOTATATE data matrix size: 128×128×79 and 2-bed PSMA data matrix size: 128×128×158, both with voxel size in mm: 4.8×4.8×4.8) with 6 subsets and 16 iterations using in-house software (*37*). No post-processing filter was applied. Scatter correction used a triple energy window method, while the depth-dependent attenuation correction used the standard CT-to-density calibration curve. The point spread function for depth-dependent collimator-detector response modeling was simulated with Monte Carlo (*38*) using a point source in air and fitted with Gaussian curves.

## Evaluation

SPECT image quality was evaluated for four distinct reconstruction methods: 1) Full reconstruction using all 120 measured projections (full recon). 2) Partial reconstruction using a certain DF of the measured projections (partial recon). 3) Linear interpolation reconstruction, where a certain DF of projections were measured, and the remaining projections were generated through linear interpolation (LinInt recon). 4) NeRF reconstruction, where a certain DF of projections were measured, and the remaining were MLP-predicted synthetic projections (NeRF recon).

We quantified reconstruction performance using multiple evaluation metrics, including Normalized Root Mean Squared Difference (NRMSD), Activity Recovery (AR), AR to Noise Ratio (ARNR), Contrast-to-Noise-Ratio (CNR) and relative CNR (RCNR). In the phantom study, the uniform 'warm' region served as the background (BKG). For the clinical patient study, we selected a homogeneous region within the lung as the BKG. The noise level was calculated as the standard deviation of voxel counts within this BKG, denoted as $\text{STD}_{\text{BKG}}$. These evaluations provide an assessment of the synthesized projection and reconstructed image compared to a reference image: the true activity map for phantom data and the OSEM reconstruction using all 120 measured projections (i.e., full recon) for patient data. Definitions of the above metrics are given as follows:

$$\text{NRMSD} = \frac{\sqrt{\frac{1}{n_p}\sum_{j=1}^{n_p}(\hat{x}_j - x_j)^2}}{\sqrt{\frac{1}{n_p}\sum_{j=1}^{n_p} x_j^2}}$$

$$\text{AR} = \frac{\text{mean counts of reconstruction within VOI}}{\text{mean counts of true activity within VOI}}, \quad \text{ARNR} = \frac{\text{AR}}{\text{STD}_{\text{BKG}}}$$

$$\text{CNR} = \frac{(\text{mean of VOI} - \text{mean of BKG})}{\text{STD}_{\text{BKG}}}, \quad \text{RCNR} = \frac{\text{CNR}_{\text{sparse view recon}}}{\text{CNR}_{\text{full recon}}} \times 100\%$$

where $n_p$ is the total number of voxels within the VOI, including lesions and relevant organs. Subscript $j$, i.e., $x_j$, denotes the $j$th voxel in the image. The reference image and the reconstructed image are denoted by $x$ and $\hat{x}$, respectively.

# RESULTS

## Synthesized Projections

Table 1 compares the performance of linearly interpolated projections against NeRF-synthesized projections, summarizing the NRMSD values across various DFs for phantom studies and patient studies. The results consistently demonstrate that the NeRF-synthesized projections outperform linearly interpolated projections, exhibiting lower NRMSD values in both phantom and patient studies.

**Table 1**. NRMSD (relative to measured projections) comparisons between NeRF-synthesized projections and linearly interpolated projections across different DFs for phantom studies and patient studies (average across 11 DOTATATE studies and 6 PSMA studies).

|        | Phantom Study | | Patient Study | | | |
|        | | | DOTATATE | | PSMA | |
|        | **NeRF** | **Linear** | **NeRF** | **Linear** | **NeRF** | **Linear** |
|---|---|---|---|---|---|---|
| **DF=2** | **5.9%** | 9.0%  | **16.9%** | 23.4% | **17.5%** | 24.6% |
| **DF=4** | **6.2%** | 9.5%  | **17.5%** | 25.5% | **18.4%** | 27.4% |
| **DF=8** | **7.5%** | 11.1% | **18.8%** | 30.4% | **23.7%** | 34.1% |

Visually, NeRF-synthesized projections appear smoother than their measured counterparts. Fig. 2 displays the measured (Fig. 2a) and synthesized projections (Figs. 2b and 2c) for a representative PSMA patient. Close examination of the intensity profiles across the lacrimals reveals notable differences: the NeRF-synthesized projection exhibits two peaks (corresponding to high uptake in left and right lacrimals as expected with PSMA), more closely aligning with the pattern observed in the measured projection, while the linearly interpolated projection presents four peaks due to angular interpolation.

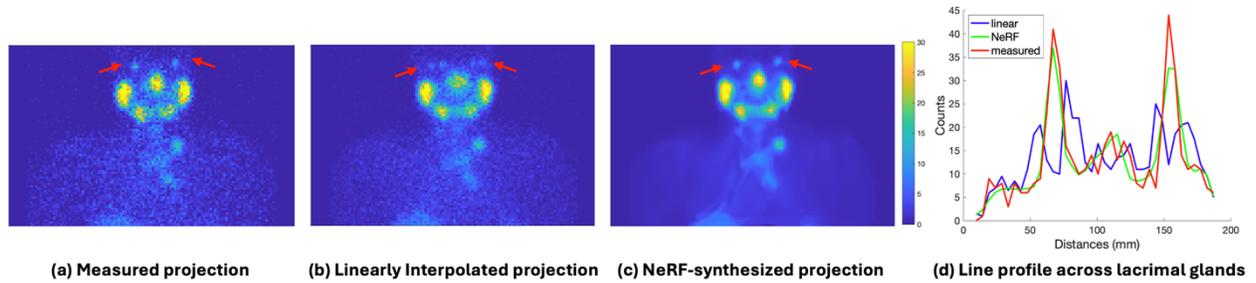

**Figure 2.** Comparison of measured and synthesized projections for a patient following Lu-177-PSMA therapy. (a), (b), and (c) show measured projection, linearly interpolated projection, and NeRF-synthesized projection, respectively. The images and profile comparison across lacrimal glands show two hot spots/peaks in the NeRF synthesized projection (green line) corresponding to left and right lacrimals, closely resembling the profile of the measured projection (red line), whereas the corresponding results for the linear interpolation shows 4 peaks due to distortions.

**Phantom Reconstruction Results**

Consider the DF=4 scenario as an illustrative case. Fig. 3 compares four reconstructions with the true activity map. Although each reconstruction method exhibits structural similarities with the true activity, the partial recon is noticeably noisier than its counterparts. Quantitative comparisons, presented in Fig. 4, plot AR to noise curves for various spheres at DF=2,4 and 8. Clearly, the NeRF recon outperforms both the partial recon and LinInt recon, delivering results that most closely parallel the full recon through various numbers of iterations of the OSEM algorithm. Note that even for the full reconstruction, AR is degraded (AR < 1) because of partial volume effects (*39*).

Moreover, the noise level in all sparse-view reconstructions increases as the DF becomes larger. But the NeRF reconstruction consistently achieved highest activity recoveries for all six lesions at the same noise level. At DF=8, detailed in Fig. 4(c), the partial reconstruction attained higher activity recovery for small lesions, at the expense of substantially increased noise level, while the NeRF reconstruction remains superior for larger lesions. For all sizes of lesions and DFs, the NeRF recon matched the activity recovery of the LinInt recon while maintaining a significantly lower noise level.

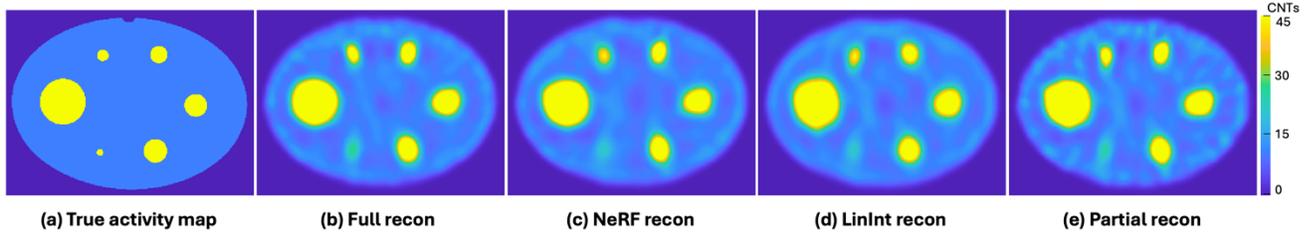

**Figure 3**. Visual comparison of (a) phantom true activity, (b) full recon, and (c) NeRF recon, (d) LinInt recon, (e) partial recon for DF=4. All images are in the same color scale.

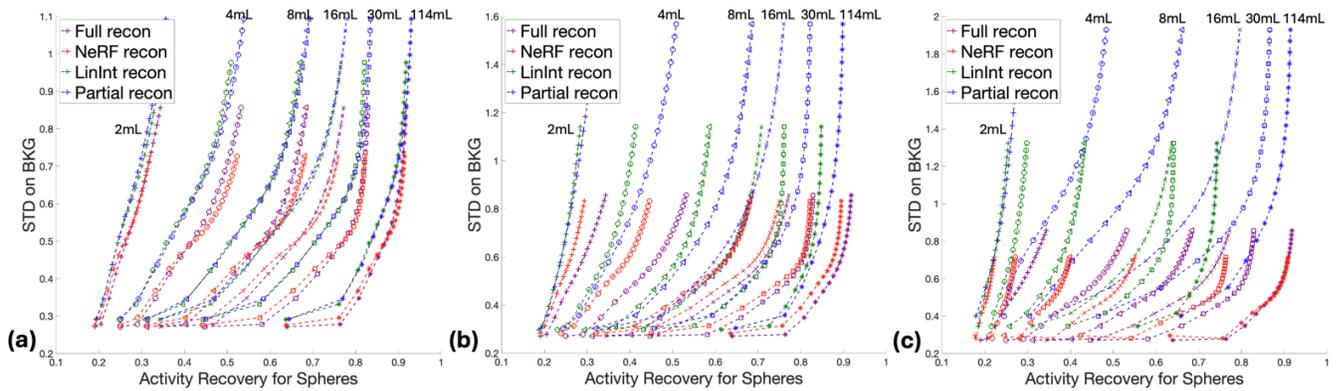

**Figure 4**. AR to noise curves for sphere volumes ranging from 2 to 114 mL for the full recon and across DFs of 2, 4, and 8 (a to c). Distinct markers are consistently used to represent each sphere volume across all subfigures. The comparison illustrates the variations in AR and noise levels across four reconstruction methods: full recon, NeRF recon, LinInt recon, and partial recon, for different sphere sizes.

**Patient Reconstruction Results**

Figs. 5 and 6 show the coronal Maximum Intensity Projections (MIPs) of an example patient image following DOTATATE and PSMA therapy, respectively, derived from four different reconstruction methods at various DFs. In both studies, the LinInt recons exhibit noticeable artifacts due to angular interpolation, more pronounced at higher DFs. This effect is particularly evident in the PSMA study for organs like the lacrimal, parotid, and submandibular glands at DF=4 and 8, substantially affecting the structural clarity of the SPECT images. Conversely, partial recons became noisier with increasing DFs, making it challenging to discern small hot spots from the background. However, the NeRF recons retained a more accurate representation of activity distribution, closely resembling the full reconstructions, while maintaining a balanced noise level.

Quantitatively, the NeRF recon yielded the highest average RCNR in the DOTATATE study, as shown in Table 2, for both lesion and kidney VOIs across all DFs. Similarly, in the PSMA study, the NeRF reconstruction had higher average RCNR for all VOIs, as shown in Table 3, across all DFs. The limitation of LinInt recon is particularly evident in the lacrimal glands, which are of very small volume (~ 0.4 mL) and exhibit exceptionally low RCNR values.

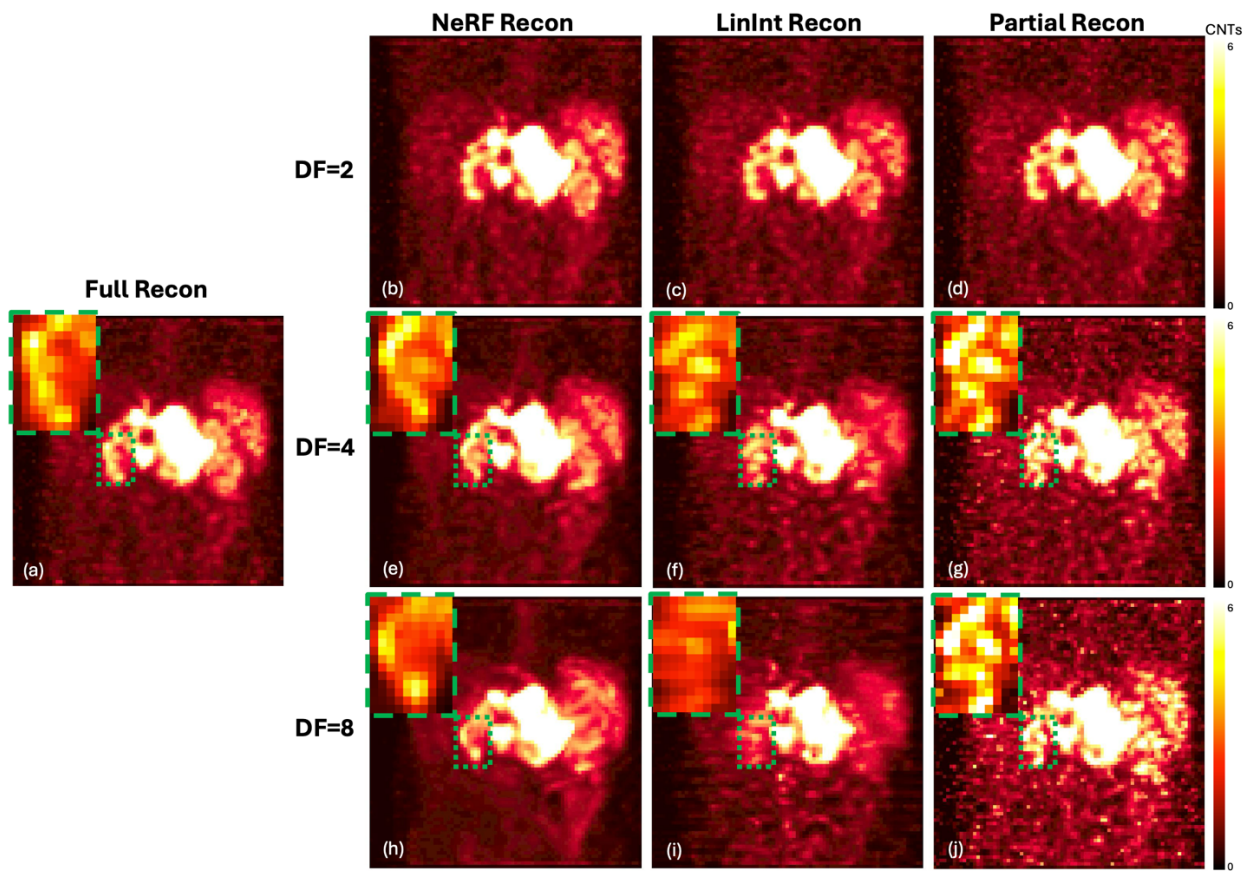

**Figure 5**. Coronal MIPs of SPECT reconstructions corresponding to a DOTATATE patient study using four reconstruction methods (columns) and three DFs (rows). Images are displayed with gamma correction with enhanced contrast levels to emphasize the blurring artifacts present in the LinInt recon and the noise present in the partial recon, especially visible at higher DFs.

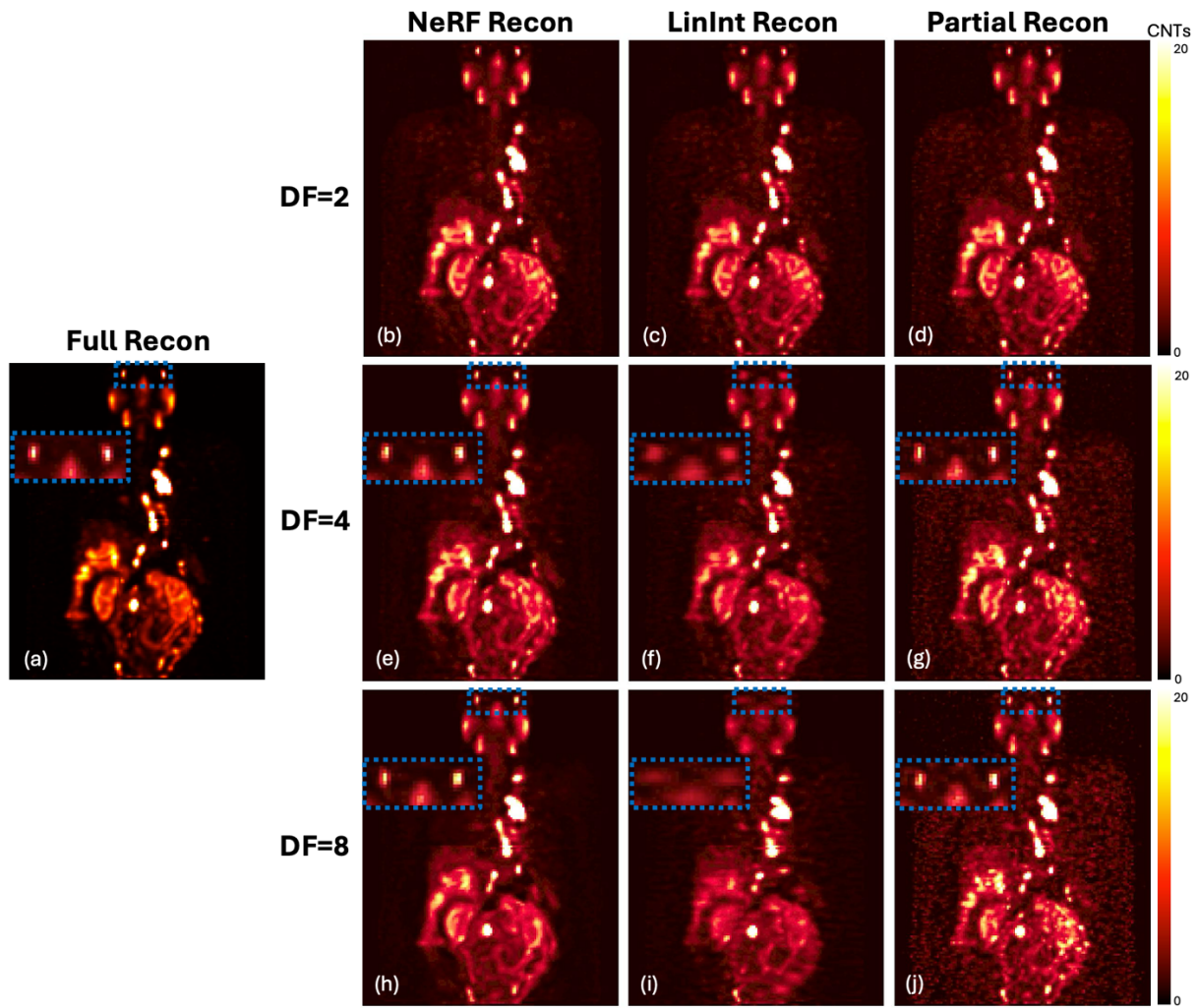

**Figure 6**. Coronal MIPs of SPECT reconstructions corresponding to a PSMA patient study using four reconstruction methods (columns) and three DFs (rows). Images are displayed with gamma correction with enhanced contrast levels to emphasize the blurring artifacts present in the LinInt recon and the noise present in the partial recon, especially visible at higher DFs.

Table 2. Average RCNR values of the NeRF recon, the LinInt recon, and the partial recon across all eleven DOTATATE patient studies, benchmarked against the full recon, whose RCNR is standardized at 100%.

|  | DF=2 | | | DF=4 | | | DF=8 | | |
|---|---|---|---|---|---|---|---|---|---|
|  | **NeRF Recon** | **LinInt Recon** | **Partial Recon** | **NeRF Recon** | **LinInt Recon** | **Partial Recon** | **NeRF Recon** | **LinInt Recon** | **Partial Recon** |
| **Lesion** | **88.6%** | 82.5% | 82.7% | **87.9%** | 68.7% | 68.7% | **73.5%** | 43.9% | 48.2% |
| **Kidney** | **92.6%** | 85.8% | 84.5% | **88.0%** | 73.1% | 67.0% | **76.5%** | 51.3% | 48.8% |

Table 3. Average RCNR values of the NeRF recon, the LinInt recon, and the partial recon across all six PSMA patient studies, benchmarked against the full recon, whose RCNR is standardized at 100%.

|  | DF=2 | | | DF=4 | | | DF=8 | | |
|---|---|---|---|---|---|---|---|---|---|
|  | **NeRF Recon** | **LinInt Recon** | **Partial Recon** | **NeRF Recon** | **LinInt Recon** | **Partial Recon** | **NeRF Recon** | **LinInt Recon** | **Partial Recon** |
| **Lesion** | **83.8%** | 79.8% | 80.7% | **78.4%** | 70.7% | 68.5% | **65.7%** | 55.7% | 54.9% |
| **All Organ ROIs** | **84.7%** | 75.7% | 80.9% | **78.4%** | 56.9% | 67.3% | **63.2%** | 31.0% | 50.8% |
| **Kidney** | **84.8%** | 79.9% | 80.3% | **80.1%** | 69.6% | 67.6% | **65.8%** | 44.2% | 51.3% |
| **Lacrimal** | **83.6%** | 63.6% | 80.4% | **77.5%** | 29.9% | 68.6% | **57.2%** | 10.2% | 47.9% |
| **Parotid** | **84.5%** | 79.3% | 80.9% | **79.1%** | 63.4% | 66.0% | **67.6%** | 34.7% | 52.0% |
| **Submandibular** | **85.6%** | 79.7% | 81.8% | **77.1%** | 64.0% | 66.9% | **62.2%** | 34.6% | 51.7% |

## DISCUSSION

The field of machine learning, particularly in the domain of deep learning (DL), is rapidly growing. Compared to other medical imaging modalities, DL applications to SPECT imaging are limited, perhaps due to the challenges of low-count scenarios of gamma-camera imaging. Previous works have demonstrated the effectiveness of using DL to generate missing SPECT projections views with convolutional neural networks, particularly, U-Net (26). However, the data-intensive nature of supervised DL makes it less feasible for SPECT imaging, where datasets are usually limited, e.g., for our study, only tens of patient data are available, and there it would be difficult to obtain hundreds or thousands of patient datasets for applying supervised learning methods. Furthermore, the change of camera-specific parameters, for example, the crystal thickness of gamma-cameras and body contour orbits, may also influence the performance of supervised learning approaches. Unlike supervised learning, self-supervised learning methods derive insights directly from the current image itself without the need for labeled datasets, making them inherently adaptable and robust to variations in testing conditions. Thus, this paper focused on a self-supervised learning method.

With evaluation both on phantoms that covered clinically relevant conditions and patients who underwent Lu-177 DOTATATE and PSMA therapy in our clinic, we have demonstrated that our NeRF recon, based on self-supervised coordinate-based learning, effectively compensates for image quality degradation under scenarios of sparse view acquisition. Considering both the reduction in acquisition time and quantitative accuracy/noise, a DF=4, appears to be a good compromise. In the patient studies, at a DF of 4, the NeRF recon achieved CNRs of ~ 80% and higher for all organs and lesions while the other sparse view methods achieved only ~ 60 to 70% relative to the full reconstruction (Table 2, 3). Despite these promising outcomes, we observed reduced activity recovery in the NeRF recon for smaller spheres (<=4 mL) in the phantom reconstruction at higher DFs, compared to the other three reconstructions (Fig. 4). This limitation could arise from the neural network's tendency to smooth over areas in low-count SPECT images due to high noise levels, leading to averaged voxel values from high noise variances. Despite the minor loss in recovery (also observed in the LinInt recons), the NeRF recons show clearly improved CNR compared with the partial recons. Additionally, at a high DF of 8, the MLP faced challenge in accurately learning the representation of the continuous measurement field due to a substantial reduction in training data, particularly impacting finer textures that fluctuate in the measured projection domains (*30*), contributing to reduced activity recovery in small lesions. Future research could explore the integration of variational inference or generative models to diversify the sampling process, potentially mitigating this smoothing effect and enhancing the model's fidelity in capturing fine details.

In a previous study investigating DL to synthesize missing projection, Monte Carlo (MC) based reconstruction was used (*24*). Although the attenuation, scatter, and collimator-detector response can be included simultaneously and accurately in the MC-based forward projection, this requires the simulation of a large number of photon histories, which is computationally expensive and therefore is less practical for routine clinical application. Instead, we used a reconstruction protocol similar to what is used in the clinic: a publicly available linear forward-backward system model with triple energy window scatter correction (*40*), which is a widely accepted and practical approach for Lu-177 SPECT reconstruction.

The idea of NeRF was to render photorealistic novel views of scenes with complicated geometries and appearances by representing a scene as a continuous function that outputs the radiance emitted in the coordinate space. To learn the continuous representation, a MLP is trained by inputting the coordinate of the scene and the training targets are the three-channel RGB colors. In this work, we conducted a similar training process where the targets were defined as single-channel SPECT projection counts. Moreover, the nature of coordinate-based learning works on the projection domain and hence is not restricted to a specific image reconstruction method but is compatible with many methods including those based on model-based image reconstruction (MBIR) (*20*) or other methods such as plug-and-play (*41*) approaches. MBIR methods often handle a complete set of projection views but with fewer counts per view. Such methods can improve image quality and reduce noise by incorporating appropriate regularizers and priors; however, choosing the optimal regularizers and regularization parameters remains a challenge. In contrast, our method is tuning-free, as evident from the good performance in two different therapies where the activity distribution in the body is substantially different.

Although our research was initially focused on Lu-177 SPECT imaging, we expect that our coordinates learning-based self-supervised method could be adapted for use in other low-count applications. This includes pure $\beta^-$-emitters, like Y-90, characterized by a low yield of bremsstrahlung photons for SPECT imaging (*42*), and $\alpha$-emitters, like Ac-225 that have low gamma-ray yields (*4*). Both present inherent low-count imaging challenges that could potentially benefit from our approach. Furthermore, our method, which allows for skipping projection views could benefit diagnostic SPECT imaging by enabling administration of lower activities, therefore supporting low-dose SPECT protocols that reduce radiation exposure to patients with minimal compromise to image quality.

# CONCLUSION

This study addresses the challenge of extended SPECT imaging durations under low-count conditions, as encountered in Lu-177 SPECT imaging, by developing a self-supervised coordinate learning approach that efficiently synthesizes skipped SPECT projection views without separate training data. The proposed method enables a significant reduction in SPECT acquisition time by allowing for skipping projection views and using a MLP to synthesize skipped projections, while preserving image quality, as indicated by improved NRMSD in projections, and ARNR and RCNR in reconstructions compared with other methods for sparse acquisitions. Unlike deep learning-based approach, this self-supervised method addresses the challenge of limited training data availability commonly encountered in clinical settings. The feasibility for reduction in acquisition time demonstrated in this work is particularly relevant for imaging under low-count conditions and for protocols that require multiple-bed positions.

# DISCLOSURES

This research was supported by R01 EB022075 awarded by the National Institute of Biomedical Imaging and Bioengineering (NIBIB) and R01 CA240706 awarded by the National Cancer Institute (NCI), NIH. Yuni Dewaraja is a consultant for MIM Software Inc. No other potential conflicts of interest relevant to this article exist.